\documentclass[journal]{IEEEtran} % use the `journal` option for ITherm conference style
\IEEEoverridecommandlockouts
\usepackage{cite}
\usepackage{amsmath,amssymb,amsfonts,bm}
\usepackage{algorithmic}
\usepackage{graphicx}
\usepackage{textcomp}
\usepackage{xcolor}
\usepackage{fancybox}
\usepackage{lipsum}
\usepackage{amsmath}
\usepackage{graphicx}
\usepackage[caption=false,font=footnotesize]{subfig} % IEEEtran-friendly
\usepackage{stfloats} % allow two-column floats at the bottom ([!b])

\def\BibTeX{{\rm B\kern-.05em{\sc i\kern-.025em b}\kern-.08em
    T\kern-.1667em\lower.7ex\hbox{E}\kern-.125emX}}
\begin{document}

\title{CLEAR: A Closed-Form Minimal-Sensor TDOA/FDOA Estimator for Moving-Source IoT Localization}
% delete or comment-out the following line before submission
\author{%
  Mohammad Kazzazi\textsuperscript{*},~Mohammad Morsali\textsuperscript{*},~and Rouhollah Amiri
  \thanks{\textsuperscript{*}These authors contributed equally to this work. The authors are with Department of Electrical Engineering, Sharif University of Technology, Tehran, Iran (Email: smohammad.ok@ee.sharif.edu, mohammad.morseli@ee.sharif.edu, amiri@sharif.edu).}%
}

\maketitle

\begin{abstract}
This paper presents CLEAR—a Closed-form Localization Estimator with A Reduced sensor network. The proposed method is a computationally efficient, two-stage estimator that fuses Time-Difference-of-Arrival (TDOA) and Frequency-Difference-of-Arrival (FDOA) measurements with a minimal number of sensors. CLEAR localizes a moving source in $N$-dimensional space using only $N{+}1$ sensors, achieving the theoretical minimum sensor count. The first stage introduces auxiliary range and range-rate parameters to construct a set of pseudo-linear equations, solved via weighted least squares. An algebraic elimination using Sylvester’s resultant then reduces the problem to a quartic equation, yielding closed-form estimates for the nuisance variables. A second, lightweight linear refinement stage is applied to mitigate residual bias. Under mild Gaussian noise assumptions, the estimator's position and velocity estimates are statistically efficient, closely approaching the Cramér–Rao Lower Bound (CRLB). Extensive Monte Carlo simulations in 2-D and 3-D scenarios demonstrate CRLB-level accuracy and consistent performance gains over representative two-stage and iterative baselines, confirming the method's high suitability for power-constrained, distributed Internet of Things (IoT) applications such as UAV tracking and smart transportation.
\end{abstract}


\section{Introduction}
\IEEEPARstart{I}{n} modern wireless systems and the Internet of Things (IoT), the accurate localization of mobile sources is a foundational capability, with critical applications in autonomous navigation for smart transportation, Unmanned Aerial Vehicle (UAV) tracking, and emergency response. In these contexts, joint processing of Time-Difference-of-Arrival (TDOA) and Frequency-Difference-of-Arrival (FDOA) measurements has been established as a superior alternative to TDOA-only approaches, offering enhanced accuracy and robustness. By fusing time-delay and Doppler-shift information, TDOA/FDOA fusion mitigates performance degradation in challenging signal environments. This advantage is grounded in estimation theory; for instance, the seminal maximum-likelihood analysis by Stoica and Li demonstrated that incorporating Doppler shifts with time delays fundamentally lowers the error bounds for localization, providing a theoretical basis for the performance gains of fused measurements \cite{Stoica2006, Pine2021}.

\subsection{Related Works}
Recent research has significantly advanced TDOA/FDOA localization by addressing various practical and non-ideal conditions. A prominent challenge is the presence of sensor position uncertainty. To mitigate this, robust estimators based on constrained least squares have been developed; for instance, the work by Yu et al. \cite{YuHuangGao2012} employs a total least-squares approach to counteract array-calibration errors. Beyond sensor inaccuracies, environmental factors are also critical. Ramezani, for example, extended passive localization to challenging underwater scenarios by explicitly modeling sound-speed gradients, demonstrating the superior tracking performance of joint TDOA/FDOA data in such dispersed media.

The impact of sensor–target geometry has also been a key focus. Pine and Cheney \cite{Pine2021} provided a rigorous geometric analysis for the far-field case, clarifying how geometric configurations affect dilution of precision and characterizing the feasibility of TDOA/FDOA measurements. In parallel, methods offering guaranteed performance bounds have emerged. Zhou and Song, for example, applied interval analysis to derive rigorous error bounds on the localization solution. For handling data outliers, which are common in real-world deployments, Cameron and Bates \cite{CameronBates2017} demonstrated a stable approach by parameterizing the problem as a polynomial system and solving it with homotopy continuation, augmented by RANSAC filtering. Collectively, these advances underscore a consistent research thrust: the development of stable and robust TDOA/FDOA algorithms capable of performing reliably under practical constraints such as sensor noise, environmental biases, and data outliers.

In addition, researchers have been increasingly paying attention to low-complexity integration of TDOA/FDOA with IoT sensing scenarios. Traditional methods involve treating TDOA and FDOA estimation as individual steps, which can result in a heavy computational burden and error sensitivity  \cite{Stoica2006}. To tackle this issue, Zhang et al. introduce a non-coherent passive sensing system that avoids individual time-delay and Doppler estimation and directly pulls out target range and velocity through a joint transform-domain examination  \cite{Zhang2024}. The method raises detection precision while bypassing independent TDOA/FDOA estimation with heavy complexity \cite{Zhang2024}. Likewise, Sun \emph{et al.} introduce a location-robust, computationally efficient positioning method for IoT devices, achieving near-optimal accuracy with minimal overhead  \cite{Sun2022}. In underwater IoT settings, Kim \emph{et al.} address hybrid FDOA/TDOA localization with additional uncertainties: their method assumes each sensor’s position error is bounded and the source’s carrier frequency may drift, yet guarantees reliable localization by using prior speed limits without requiring prior noise statistics \cite{Kim2025}. At the system level, SPIN shows how downlink TDOA/FDOA on non-terrestrial networks (NTN) synchronization signals can yield joint position/velocity estimates that meet CRLB accuracy and enable uplink synchronization with substantial battery-life savings for power-constrained IoT UEs  \cite{Chandrika2023}. Complementarily, robust IoT TDOA methods that jointly estimate source position and NLOS bias improve resilience in urban deployments \cite{Wang2019}; joint target localization and sensor self-calibration of positions and synchronization via sequential range/angle observations achieves CRLB-level accuracy with closed-form and SDR variants  \cite{Jia2025}; and altitude-constrained fusion of TDOA, FDOA, and differential Doppler rate yields efficient iterative WLS updates with derived CRLBs for known-altitude sources  \cite{Ma2022}. Finally, for range- and frequency-type measurements in range-dependent media, underwater TOA/FOA formulations that incorporate isogradient sound-speed profiles are necessary to approach CRLB performance  \cite{Qin2025}. 

\subsection{Motivation}

The joint TDOA/FDOA localization problem has been extensively studied for both stationary and mobile emitters. Early algebraic solutions, such as the closed-form solution for a mobile source by Ho and Xu \cite{HoXu2004}, demonstrated the significant potential of fusing these measurements. Traditional approaches often rely on a two-step Weighted Least Squares (WLS) process, which can achieve high accuracy but typically demands a surplus of sensors and a carefully chosen initial guess. Alternatively, nonlinear solvers like Taylor-series expansion or quasi-Newton methods are susceptible to local minima without a sufficiently accurate initialization.

To circumvent the issue of local minima, convex relaxation techniques such as Semidefinite Programming (SDP) have been employed \cite{WangLiAnsari2013}; however, these methods often incur substantial computational complexity, limiting their use in resource-constrained applications. This has motivated the development of closed-form estimators, which eliminate iterative hazards by providing explicit solutions. The seminal Chan–Ho algorithm, for instance, offers a direct solution for TDOA-only localization \cite{ChanHo1994}, and its successors often construct pseudo-linear equations to estimate source coordinates without iteration \cite{ChenEtAl2023}.

A fundamental limitation persists, however: most existing closed-form schemes require more measurements than the theoretical minimum to fully linearize the problem. For example, the Chan–Ho TDOA estimator requires one extra sensor beyond the minimum in 3-D space \cite{ChanHo1994}. While other methods, such as the two-stage solver by Wang and Li that incorporates Doppler-rate information \cite{WangLi2021}, achieve Cramér–Rao Lower Bound (CRLB) level accuracy, they do not operate at the minimum sensor threshold. Recent geometric analyses have further clarified the theoretical limits of TDOA/FDOA measurements \cite{Pine2021}, yet a critical gap remains.

To the best of our knowledge, no existing closed-form estimator can localize a moving source's position and velocity with the theoretical minimum number of sensors. This paper aims to fill this gap by proposing a computationally efficient, closed-form solution that requires only $N+1$ sensors for $N$-dimensional localization.

\subsection{Contributions}
Achieving accurate localization with the minimum number of sensors has been a focal point of recent work. Theoretically, in $ N$-dimensional space, at least $N$ independent range-difference measurements are needed to localize an unknown source. Traditional solutions often exceed this minimum. Amiri \emph{et al.} addressed this gap for TDOA-only localization, showing that a two-stage WLS approach can attain the CRLB using exactly $N{+}1$ sensors (e.g., three sensors in 2-D) by introducing an auxiliary range parameter \cite{AmiriBehniaNoroozi2018}. Similarly, Noroozi \emph{et al.} derived a closed-form solution for multistatic radar that localizes a target with the fewest transmit/receive nodes (e.g., one transmitter and three receivers in 3-D) \cite{NorooziEtAl2020}. These minimal-sensor solutions, however, assume static targets or omit velocity estimation. In practice, if fewer than $N{+}1$ sensors are available at an instant, one must leverage target motion over time (e.g., multiple epochs) to accumulate information; motion-assisted TDOA/FDOA strategies can restore observability with sub-minimal arrays at the expense of temporal association and latency.

In this work, we introduce CLEAR, a two-stage algebraic estimator that fuses TDOA and FDOA with only $N{+}1$ sensors in $N$-D. Stage~1 of CLEAR introduces range and range-rate nuisance parameters, solves a pseudo-linear WLS, and uses a Sylvester resultant to reduce the system to a quartic whose real root yields $(v,\dot{v})$ in closed form. Stage~2 applies a lightweight linear refinement that corrects residual bias. The resulting non-iterative pipeline jointly recovers position and velocity and, under mild Gaussian noise, attains CRLB-level performance. Unlike prior closed-form schemes that ignore Doppler or require extra sensors/snapshots, CLEAR preserves observability at the theoretical minimum—aligned with IoT constraints on sensor count, compute, and latency.
The main contributions of this work are as follows:
\begin{itemize}
  \item \textit{Pseudo‐linearization}: we introduce range and range‐rate nuisance parameters to form pseudo‐linear equations solvable via a single WLS estimator.
  \item \textit{Algebraic elimination}: we apply Sylvester’s resultant to eliminate nuisance parameters, reducing the problem to a single quartic whose real root yields both range and range‐rate in closed form.
  \item \textit{Linear refinement}: a fast second stage corrects residual errors and produces position and velocity estimates that attain the CRLB under mild Gaussian noise.
  \item \textit{Minimal‐sensor operation}: we prove that the estimator requires only $N+1$ sensors in $N$-D space and outperforms existing two‐stage and iterative methods in low‐sensor or high‐noise regimes.
  \item \textit{Comprehensive validation}: extensive Monte Carlo simulations in 2‐D and 3‐D scenarios demonstrate CRLB‐level accuracy and robustness across challenging geometries.
\end{itemize}

\subsection{Outline}
The remainder of this paper is organized as follows.
Section~\ref{sec:problem_statement} formulates the measurement model.
Section~\ref{sec:proposed_method} presents the details of the proposed two‐stage algebraic estimator.
Section~\ref{sec:performance_analysis} analyzes performance and CRLB attainment.
Section~\ref{sec:simulation_results} presents numerical results,
and finally, Section~\ref{sec:conclusion} concludes the paper.

\section{Problem Statement}\label{sec:problem_statement}
Consider the problem of single source localization in $N$-dimensional $(N = 2,3)$ space using a sensor network consisting of $M+1$ sensors, the positions of which are denoted by $\mathbf{s}_i$, $i = 0, 1, \dots, M$. We aim to locate a source, whose true position and velocity are unknown and denoted by $\mathbf{u}^o$ and $\dot{\mathbf{u}}^o$, using a set of TDOA and FDOA measurements calculated in the receivers. Without loss of generality, we select $\mathbf{s}_0$ as the reference sensor. The received signals in the reference and other sensors are employed to extract $M$ TDOA and FDOA measurements by local processing. Subsequently, the sensors' measurements are collected in a fusion center to locate the source in a centralized manner.

The true TDOA in the $i$-th sensor, after multiplying by the wave propagation speed, is given by
\begin{equation}
r_i^o = \|\mathbf{u}^o - \mathbf{s}_i\| - \|\mathbf{u}^o - \mathbf{s}_0\|
, \quad i = 1, \dots, M.
\tag{1}
\end{equation}
We use the terms 'range' and 'delay' interchangeably, as they differ only in a constant wave propagation speed coefficient.

The observed TDOA is represented by $r_i = r_i^o + \Delta r_i$ due to measurement noise.
Collecting all TDOA measurements yields in matrix form:
\begin{equation}
\mathbf{r} = \mathbf{r}^o + \Delta\mathbf{r}
\tag{2}
\end{equation}
where $\mathbf{r}^o = [r_1^o, \dots, r_M^o]^T$ and $\mathbf{r}, \Delta\mathbf{r}$ are defined similarly. The noise vector $\Delta \mathbf{r}$ is modeled as a zero-mean Gaussian random vector with covariance $\mathbf{C}_{TDOA}$.

Furthermore, the FDOA is a measure of the Doppler shift caused by the relative velocity between the moving source and the sensors. The true FDOA measurement at the $i$-th sensor is given by the difference in the Doppler shifts between sensor $i$ and the reference sensor, which can be written as:
\begin{align}
  \dot{r}_i^o &= \frac{(\mathbf{u}^o - \mathbf{s}_i)^{T}(\dot{\mathbf{u}}^o - \dot{\mathbf{s}}_i)}
                 {\|\mathbf{u}^o - \mathbf{s}_i\|}
         - \frac{(\mathbf{u}^o - \mathbf{s}_0)^{T}(\dot{\mathbf{u}}^o - \dot{\mathbf{s}}_0)}
                  {\|\mathbf{u}^o - \mathbf{s}_0\|}, 
  \notag\\
  i &= 1,\dots,M.
  \tag{3}
\end{align}

As with TDOA, the FDOA measurements are corrupted by noise. The observed FDOA at the $i$-th sensor is:
\begin{equation}
\dot{r}_i = \dot{r}_i^o + \Delta\dot{r}_i,
\tag{4}
\end{equation}
where $ \dot{r}_i $ is the observed FDOA measurement, and $ \Delta \dot{r}_i $ represents the measurement noise, which is modeled as a zero-mean Gaussian random.
Similar to the TDOA case, we can collect the FDOA measurements in matrix form.
Collecting all FDOA measurements yields in matrix form
\begin{equation}
\dot{\mathbf{r}} = \dot{\mathbf{r}}^o + \Delta\mathbf{\dot{r}}
\tag{5}
\end{equation}
where $\dot{\mathbf{r}} = [\dot{r}_1^o, \dots, \dot{r}_M^o]^T$ and $\dot{\mathbf{r}}, \Delta \dot{\mathbf{r}}$ are defined similarly. The noise vector $\Delta \dot{\mathbf{r}}$ is modeled as a zero-mean Gaussian random vector with covariance $\mathbf{C}_{FDOA}$.

For ease of notation, we denote the total measurements 
consisting of the two sets of TDOA and FDOA measurements by $\mathbf{m} = [\mathbf{r}^T, \mathbf{\dot{r}}^T]^T$ and $\mathbf{m^o} = [\mathbf{r}^{oT}, \mathbf{\dot{r}}^{oT}]^T$ Therefore, 
$\mathbf{\Delta m} = [\Delta\mathbf{r}^T, \Delta\mathbf{\dot{r}}^T]^T$
represents the corresponding stacked noise vector of TDOA and FDOA measurements, 
which is a Gaussian random vector with zero mean and covariance matrix
\begin{equation}
\mathbf{Q}_m = \mathbb{E}[\Delta \mathbf{m} \Delta \mathbf{m}^T]= \begin{bmatrix}
    \mathbf{C}_{\text{TDOA}} & \mathbf{0} \\
    \mathbf{0} & \mathbf{C}_{\text{FDOA}}
\end{bmatrix}
.
\tag{6}\end{equation} 

Under the Gaussian noise assumption, the ML estimation 
of the source position and velocity vector 
$\boldsymbol{\theta} = [\mathbf{u}^{oT}, \mathbf{\dot{u}}^{oT}]^T$
is given by: 
\begin{equation}
\min_{\mathbf{\theta}} \left( \mathbf{m} - \mathbf{m^o}(\boldsymbol{\theta}) \right)^T \mathbf{Q}_m^{-1} \left( \mathbf{m} - \mathbf{m^o}(\boldsymbol{\theta}) \right) ,
\tag{7}
\end{equation}

In the next section, we propose a closed-form solution for this problem, which can achieve the CRLB performance under mild noise conditions.

\section{Proposed Method}\label{sec:proposed_method}
In this section, we propose a two-stage estimator for the localization problem to attain CRLB accuracy. In the first stage, a set of pseudo-linear equations is established by introducing the range nuisance parameter, which can be solved using a WLS estimator. In the second stage, the source position and velocity error terms are estimated to refine the initial solution in the first stage.

Stage 1: By rearranging (1) as $r_i^o + \|\mathbf{u}^o - \mathbf{s}_0\| = \|\mathbf{u}^o - \mathbf{s}_i\|$, squaring both sides, and simplifying, we have
\begin{equation}
{r_i^o}^2 + \|\mathbf{s}_0\|^2 - \|\mathbf{s}_i\|^2 + 2(\mathbf{s}_i - \mathbf{s}_0)^T \mathbf{u}^o + 2r_i^o v = 0
\tag{8}
\end{equation}
where $v = \|\mathbf{u}^o - \mathbf{s}_0\|$.

By replacing the true terms with their noisy values, (8) becomes
\begin{equation}
r_i^2 + \|\mathbf{s}_0\|^2 - \|\mathbf{s}_i\|^2 + 2(\mathbf{s}_i - \mathbf{s}_0)^T \mathbf{u}^o + 2r_i v \approx 2\|\mathbf{u}^o - \mathbf{s}_i\|\Delta{r_i}
\tag{9}
\end{equation}
where the second-order noise terms have been ignored.

Stacking (9) for $i = 1, \dots, M$ yields
\begin{equation}
\mathbf{h} - \mathbf{G}\mathbf{u}^o + 2\mathbf{r}v = \mathbf{B}\Delta \mathbf{r}
\tag{10}
\end{equation}
where the $i$-th element of the regressand is $[\mathbf{h}]_i = r_i^2 + \|\mathbf{s}_0\|^2 - \|\mathbf{s}_i\|^2$, the $i$-th row of the regressor is $[\mathbf{G}]_{i,:} = 2[\mathbf{s}_0 - \mathbf{s}_i]^T$, and the matrix $\mathbf{B}$ is given by
\begin{equation}
\mathbf{B} = 2 \text{diag}([\|\mathbf{u}^o - \mathbf{s}_1\|, \dots, \|\mathbf{u}^o - \mathbf{s}_M\|]^T).
\tag{11}
\end{equation}
Taking the time derivative of (10) yields
\begin{equation}
\mathbf{\dot{h}} - \mathbf{\dot{G}}\mathbf{u}^o - \mathbf{G}\mathbf{\dot{u}}^o + 2\mathbf{\dot{r}}v + 2\mathbf{r}\dot{v} = \mathbf{\dot{B}} \Delta \mathbf{r} + \mathbf{B}\Delta \mathbf{\dot{r}},
\tag{12}
\end{equation}
where the $i$-th element of the regressand is $[\mathbf{\dot{h}}]_i = 2r_i \dot{r}_i + 2\mathbf{\dot{s}}_0^T \mathbf{s}_0 - 2\mathbf{\dot{s}}_i^T \mathbf{s}_i$, the $i$-th row of the regressor is $[\mathbf{\dot{G}}]_{i,:} = 2[\mathbf{\dot{s}}_0 - \mathbf{\dot{s}}_i]^T$, and the matrix $\mathbf{\dot{B}}$ is given by
\begin{equation}
\mathbf{\dot{B}} = 2 \text{diag}\!\left(\left[(\mathbf{\dot{s}}_i - \mathbf{\dot{u}}^o)^{T} \frac{(\mathbf{s}_i - \mathbf{u}^o)}{\|\mathbf{s}_i - \mathbf{u}^o\|}\right]_{i=1}^M\right).
\tag{13}
\end{equation}
Combining (10) and (12) yields 
\begin{equation}
\mathbf{h_1} - \mathbf{G}_1\boldsymbol{\theta} + \mathbf{D}_1\boldsymbol{\phi} = \mathbf{B}_1\Delta\mathbf{m},
\tag{14}
\end{equation}
where
\[
\mathbf{G}_1 = \begin{bmatrix} \mathbf{G} & \mathbf{0} \\ \mathbf{\dot{G}} & \mathbf{G} \end{bmatrix} , \quad
\mathbf{D}_1 = \begin{bmatrix} 2\mathbf{r} & \mathbf{0} \\ 2\mathbf{\dot{r}} & 2\mathbf{r} \end{bmatrix}, \quad
\mathbf{B}_1 = \begin{bmatrix} \mathbf{B} & \mathbf{0} \\ \mathbf{\dot{B}} & \mathbf{B} \end{bmatrix},
\]
\[
\mathbf{h}_1 = \begin{bmatrix} \mathbf{h} \\ \mathbf{\dot{h}} \end{bmatrix} ,
\quad
\boldsymbol{\theta} = \begin{bmatrix} \mathbf{u}^o \\ \mathbf{\dot{u}}^o \end{bmatrix} ,
\quad
\boldsymbol{\phi} = \begin{bmatrix} v \\ \dot{v} \end{bmatrix}.
\tag{15}\]

The WLS solution of (14), can be written in terms of $\boldsymbol{\phi}^o$, as follows:
\begin{equation}
\boldsymbol{\hat{\theta}} = \left( \mathbf{G}_1^T \mathbf{W}_1 \mathbf{G}_1 \right)^{-1} \mathbf{G}_1^T \mathbf{W}_1 \left( \mathbf{h}_1 + \mathbf{D}_1 \boldsymbol{\phi} \right),
\tag{16}
\end{equation}
where $\mathbf{W}_1$ is the weighting matrix given by
\begin{equation}
\mathbf{W}_1  = \left( {\mathbf{B}}_1 \mathbf{Q_m} {\mathbf{B}}_1^T \right)^{-1}.
\tag{17}
\end{equation}
 where $\mathbf{Q_m}$ defined in (6), Note that in (16), the two nuisance parameters expressed by the vector $\boldsymbol{\phi}$ are unknown. In the following discussion, the relationships among the nuisance parameters and the source position and velocity are used to determine these two unknown parameters. By subtracting $\left[ \mathbf{s}_{0}^T, \dot{\mathbf{s}}_{0}^T \right]^T$ from both sides of (16) and introducing the new notations, (16) can be written as

\begin{equation}
\begin{cases}
    \mathbf{u} - \mathbf{s}_0 = \mathbf{b} - \mathbf{A}\boldsymbol{\phi}, \\
    \mathbf{\dot{u}} - \dot{\mathbf{s}}_0 = \dot{\mathbf{b}} - \dot{\mathbf{A}}\boldsymbol{\phi},
\end{cases}
\tag{18}
\end{equation}
where
\begin{align}
\begin{bmatrix}
\mathbf{b}^T, \dot{\mathbf{b}}^T
\end{bmatrix}^T &= \left( \mathbf{G}_1^T \mathbf{W}_1 \mathbf{G}_1 \right)^{-1} \mathbf{G}_1^T \mathbf{W}_1 \mathbf{h}_1 - \left[ \mathbf{s}_{0}^T, \dot{\mathbf{s}}_{0}^T \right]^T,\nonumber\\
\begin{bmatrix}
\mathbf{A}^T, \dot{\mathbf{A}}^T
\end{bmatrix}^T &= \left( \mathbf{G}_1^T \mathbf{W}_1 \mathbf{G}_1 \right)^{-1} \mathbf{G}_1^T \mathbf{W}_1 \mathbf{D}_1.
\tag{19}
\end{align}

Note that the two nuisance parameters, $v$ and $\dot{v}$, are related to the target position and velocity as follows
\begin{equation}
\begin{cases}
    v^2 = (\mathbf{u} - \mathbf{s}_0)^T (\mathbf{u} - \mathbf{s}_0), \\
    \dot{v}v = (\dot{\mathbf{u}} - \dot{\mathbf{s}}_0)^T (\mathbf{u} - \mathbf{s}_0).
\end{cases}
\tag{20}
\end{equation}

Multiplying both sides of the two equations in (18) by $(\mathbf{u} - \mathbf{s}_0)^T$, inserting (20) into the results and applying algebraic manipulations, eliminates $\mathbf{u}$ and $\dot{\mathbf{u}}$ from the results and yields the two following quadratic polynomial equations in terms of $\boldsymbol{\phi}$ as
\begin{equation}
\begin{cases}
\boldsymbol{\phi}^T \left( \mathbf{A}^T \mathbf{A} - \begin{bmatrix} 1 & 0 \\ 0 & 0 \end{bmatrix} \right) \boldsymbol{\phi} - 2 \mathbf{b}^T \mathbf{A} \boldsymbol{\phi} + \mathbf{b}^T \mathbf{b} = 0, \\
\boldsymbol{\phi}^T \left( \dot{\mathbf{A}}^T \mathbf{A} - \begin{bmatrix} 0 & 1 \\ 0 & 0 \end{bmatrix} \right) \boldsymbol{\phi} - \left( \dot{\mathbf{b}}^T \mathbf{A} + \mathbf{b}^T \dot{\mathbf{A}} \right) \boldsymbol{\phi} + \dot{\mathbf{b}}^T \mathbf{b} = 0.
\end{cases}
\tag{21}
\end{equation}
For better understanding, we rewrite (21) in terms of the two nuisance parameters $v$ and $\dot{v}$ as follows
\begin{equation}
\begin{cases}
f_1 : a_1 v^2 + b_1 v \dot{v} + c_1 \dot{v}^2 + d_1 v + e_1 \dot{v} + f_1 = 0, \\
f_2 : a_2 v^2 + b_2 v \dot{v} + c_2 \dot{v}^2 + d_2 v + e_2 \dot{v} + f_2 = 0,
\end{cases}
\tag{22}
\end{equation}
where
\begin{align}
a_1 &= \left[ \mathbf{A}^T \mathbf{A} \right]_{1,1} - 1, \quad a_2 = \left[ \dot{\mathbf{A}}^T \mathbf{A} \right]_{1,1},\nonumber\\
b_1 &= 2 \left[ \mathbf{A}^T \mathbf{A} \right]_{1,2}, \quad b_2 = \left[ \dot{\mathbf{A}}^T \mathbf{A} \right]_{1,2} + \left[ \dot{\mathbf{A}}^T \mathbf{A} \right]_{2,1} - 1,\nonumber\\
c_1 &= \left[ \mathbf{A}^T \mathbf{A} \right]_{2,2}, \quad c_2 = \left[ \dot{\mathbf{A}}^T \mathbf{A} \right]_{2,2},\nonumber\\
d_1 &= -2 \mathbf{b}^T \left[ \mathbf{A} \right]_{:,1}, \quad d_2 = -\dot{\mathbf{b}}^T \left[ \mathbf{A} \right]_{:,1} - \mathbf{b}^T \left[ \dot{\mathbf{A}} \right]_{:,1},\nonumber\\
e_1 &= -2 \mathbf{b}^T \left[ \mathbf{A} \right]_{:,2}, \quad e_2 = -\dot{\mathbf{b}}^T \left[ \mathbf{A} \right]_{:,2} - \mathbf{b}^T \left[ \dot{\mathbf{A}} \right]_{:,2},\nonumber\\
f_1 &= \mathbf{b}^T \mathbf{b}, \quad f_2 = \dot{\mathbf{b}}^T \mathbf{b}.
\tag{23}
\end{align}
% --- Wide resultant equation at the bottom of the page ---
\begin{figure*}[!b] % requires \usepackage{stfloats}
\hrulefill
\vspace{-1.2ex}

% If this equation is NOT physically placed after Eq. (23) in your .tex,
% uncomment the next line to force the number (24):
\setcounter{equation}{23}

\begin{equation}
\mathrm{Res}\!\left(f_1,f_2,\hat{v}\right)
=
\det\!\begin{bmatrix}
C_1 & 0 & C_2 & 0 \\
B_1 \hat{v} + E_1 & C_1 & B_2 \hat{v} + E_2 & C_2 \\
A_1 \hat{v}^2 + D_1 \hat{v} + F_1 & B_1 \hat{v} + E_1 & A_2 \hat{v}^2 + D_2 \hat{v} + F_2 & B_2 \hat{v} + E_2 \\
0 & A_1 \hat{v}^2 + D_1 \hat{v} + F_1 & 0 & A_2 \hat{v}^2 + D_2 \hat{v} + F_2
\end{bmatrix}.
\label{eq:resultant_24}
\end{equation}

\vspace{-1.2ex}
\end{figure*}
 
To solve (22) and find the common roots of $f_1$ and $f_2$, we eliminate $\dot{v}$ and form a quartic polynomial function of $v$ using the elimination method based on the resultant. The resultant of $f_1$ and $f_2$ with respect to $\dot{v}$, denoted by $\text{Res}(f_1, f_2, \dot{v})$, is defined as the determinant of the Sylvester matrix and is given by (24), shown at the bottom of the next page. Computing the determinant in (24) leads to a polynomial of degree four as follows
\begin{equation}
p_1 v^4 + p_2 v^3 + p_3 v^2 + p_4 v + p_5 = 0,
\tag{25}
\end{equation}
where $p_i$ for $i=1, \dots, 5$ can be computed by the symbolic toolbox of MATLAB \textcopyright{} and are not included in the paper due to their long terms. When the roots of (25) are found, the corresponding values for $\dot{v}$ are computed by inserting each of them into the following equation
\begin{equation}
\hat{\dot{v}} = -  
\dfrac{ 
\left|\begin{matrix}
c_1 & a_1 \\
c_2 & a_2
\end{matrix}\right|
\hat{v}^2
+ 
\left|\begin{matrix}
c_1 & d_1\\
c_2 & d_2
\end{matrix}\right|
\hat{v} 
+ 
\left|\begin{matrix}
c_1 & f_1 \\
c_2 & f_2
\end{matrix}\right|
}{
\left|\begin{matrix}
c_1 & b_1 \\
c_2 & b_2
\end{matrix}\right|
\hat{v}
+ 
\left|\begin{matrix}
c_1 & e_1 \\
c_2 & e_2
\end{matrix}\right|
}
\tag{26}
\end{equation}
It is worth noting that since the nuisance parameter $v$ is the distance to the source from the reference sensor, only positive values are acceptable for $v$. Then, the eligible nuisance parameters are substituted into (16) to find the source position and velocity estimates in the first stage.

\textit{Remark 1:} When there exists more than one solution for $\boldsymbol{\phi}$, accordingly, we have more than one solution for $\boldsymbol{\theta}$ through (16). In such a case, to eliminate this ambiguity and determine the correct solution for $\boldsymbol{\theta}$, we insert all the candidate solutions into the ML estimator given by (7) and choose the one with the smaller ML cost function.

\textit{Remark 2:} The weighting matrix $\mathbf{W}_1$ in (16) is dependent on the two unknown nuisance parameters, $v$ and $\dot{v}$, through $\mathbf{B}_1$. In order to implement the algorithm, we first set $\mathbf{W}_1 = \mathbf{Q}_m^{-1}$ to generate an initial estimate for the nuisance parameters. The initial estimates are then exploited to form the desirable weighting matrix according to (17). Using the new weighting matrix in (16) can lead to a more accurate estimate of the source position and velocity.

Stage 2: In this stage, we aim to estimate the error terms \(\Delta \mathbf{u}\) and \(\Delta \dot{\mathbf{u}}\), refining the initial estimate in the first stage, to improve the localization accuracy.

By replacing the true terms with their erroneous values as \(\mathbf{u}^o = \hat{\mathbf{u}} - \Delta \mathbf{u}\), (9) can be written as
\[
r_i^2 + \|\mathbf{s}_0\|^2 - \|\mathbf{s}_i\|^2 + 2(\mathbf{s}_i - \mathbf{s}_0)^T \hat{\mathbf{u}} + 2r_i \|\hat{\mathbf{u}} - \mathbf{s}_0\| 
\]
\[
 + 2(\mathbf{s}_0 - \mathbf{s}_i - r_i \boldsymbol{\rho}_{\hat{\mathbf{u}}, \mathbf{s}_0})^T \Delta \mathbf{u} \approx 2\|\mathbf{u}^o - \mathbf{s}_i\| \Delta \mathbf{r}_i  ,
\tag{27}\]
where \(\boldsymbol{\rho}_{\mathbf{a},\mathbf{b}} = \frac{(\mathbf{a} - \mathbf{b})}{\|\mathbf{a - b}\|}\) and we have approximated the term \(v = \|\hat{\mathbf{u}} - \mathbf{s}_0 - \Delta \mathbf{u}\|\) using first-order Taylor series expansion as
\begin{equation}
v = \|\hat{\mathbf{u}} - \mathbf{s}_0 - \Delta \mathbf{u}\| \approx \|\hat{\mathbf{u}} - \mathbf{s}_0\| - \boldsymbol{\rho}_{\hat{\mathbf{u}}, \mathbf{s}_0}^T \Delta \mathbf{u}.
\tag{28}
\end{equation}
Equation (28) can be written in matrix form as
\begin{equation}
\mathbf{d} - \mathbf{F} \Delta \mathbf{u} = \mathbf{B} \Delta \mathbf{r}
\tag{29}
\end{equation}
where \(\mathbf{B}\) is defined in (11). The \(i\)-th element of \(\mathbf{d}\) and the \(i\)-th row of \(\mathbf{F}\), respectively, are
\begin{align}
[\mathbf{d}]_i &= r_i^2 + \|\mathbf{s}_0\|^2 - \|\mathbf{s}_i\|^2 + 2(\mathbf{s}_i - \mathbf{s}_0)^T \hat{\mathbf{u}} + 2r_i \|\hat{\mathbf{u}} - \mathbf{s}_0\|,\nonumber\\
[\mathbf{F}]_{i,:} &= 2[r_i \boldsymbol{\rho}_{\hat{\mathbf{u}}, \mathbf{s}_0}^T + \mathbf{s}_i^T - \mathbf{s}_0^T].
\tag{30}
\end{align}
Taking the time derivative of (29) yields
\begin{equation}
\mathbf{\dot{d}} - \mathbf{\dot{F}} \Delta \mathbf{u}  - \mathbf{F} \Delta \mathbf{\dot{u}} = \mathbf{\dot{B}} \Delta \mathbf{r} + \mathbf{B} \Delta \mathbf{\dot{r}}
\tag{31}
\end{equation}
where \(\mathbf{\dot{B}}\) is defined in (13). The \(i\)-th element of \(\dot{\mathbf{d}}\) and the \(i\)-th row of \(\dot{\mathbf{F}}\), respectively, are 
\begin{align}
[\mathbf{\dot{d}}]_i &= 2\dot{r}_ir_i + 2\mathbf{\dot{s}}_0^T \mathbf{s}_0 - 2\mathbf{\dot{s}}_i^T \mathbf{s}_i + 2(\mathbf{\dot{s}}_i - \mathbf{\dot{s}}_0)^T \hat{\mathbf{u}} \nonumber\\
&+ 2(\mathbf{s}_i - \mathbf{s}_0)^T \hat{\dot{\mathbf{u}}} + 2\dot{r}_i \|\hat{\mathbf{u}} - \mathbf{s}_0\| + 2r_i\boldsymbol{\rho}_{\hat{\mathbf{u}}, \mathbf{s}_0}^T \left( \hat{\dot{\mathbf{u}}} - \dot{\mathbf{s}}_0 \right)\nonumber\\
[\mathbf{\dot{F}}]_{i,:} &= 2 \left( \dot{r}_i \boldsymbol{\rho}_{\hat{\mathbf{u}}, \mathbf{s}_0}^T + r_i  \dot{\boldsymbol{\rho}}_{\hat{\mathbf{u}}, \mathbf{s}_0}^T + \mathbf{\dot{s}}_i^T - \mathbf{\dot{s}}_0^T \right)
\tag{32}
\end{align}
where $\dot{\boldsymbol{\rho}}_{\mathbf{a},\mathbf{b}}$ is defined as
\begin{equation}
\dot{\boldsymbol{\rho}}_{\mathbf{a},\mathbf{b}} = \left( \mathbf{I} - \frac{(\mathbf{a} - \mathbf{b})(\mathbf{a} - \mathbf{b})^T}{\|\mathbf{a} - \mathbf{b}\|^2} \right) \frac{(\mathbf{\dot{a}} - \mathbf{\dot{b}})}{\|\mathbf{a} - \mathbf{b}\|}.
\tag{33}
\end{equation}

Combining (29) and (31) yields 
\begin{equation}
\mathbf{h}_2 - \mathbf{G}_2\Delta\boldsymbol{\theta}  = \mathbf{B}_2\Delta\mathbf{m},
\tag{34}
\end{equation}
where
\[
\mathbf{G}_2 = \begin{bmatrix} \mathbf{A} & \mathbf{0} \\ \mathbf{\dot{A}} & \mathbf{A} \end{bmatrix} , \quad
\mathbf{B}_2 = \begin{bmatrix} \mathbf{B} & \mathbf{0} \\ \mathbf{\dot{B}} & \mathbf{B} \end{bmatrix}
\]
\[
\mathbf{h}_2 = \begin{bmatrix} \mathbf{b} \\ \mathbf{\dot{b}} \end{bmatrix} ,
\quad
\Delta\boldsymbol{\theta} = \begin{bmatrix} \Delta\mathbf{u} \\ \Delta\mathbf{\dot{u}} \end{bmatrix} .
\tag{35}\]

The WLS solution of (34) is given by 
\begin{equation}
\Delta \boldsymbol{\hat{\theta}} = \left( \mathbf{G}_2^T \mathbf{W}_2 \mathbf{G}_2 \right)^{-1} \mathbf{G}_2^T \mathbf{W}_2 \mathbf{h}_2
\tag{36}
\end{equation}
where
\begin{equation}
    \mathbf{W}_2 = \left( \mathbf{B}_2 \mathbf{Q}_m \mathbf{B}_2^T \right)^{-1}.
\tag{37}
\end{equation}
After solving (36), the final source position and velocity estimate are obtained as
\begin{equation}
    \bar{\boldsymbol{\theta}} = \hat{\boldsymbol{\theta}} - \Delta \hat{\boldsymbol{\theta}}
\tag{38}
\end{equation}

\textit{Remark 3:} The minimum number of required measurements for both stages of the proposed method is $N$, which is equivalent to the existence of $N + 1$ sensors. It is noteworthy that the minimum number of required sensors for $N$-D TDOA/FDOA-based localization in general is also $N + 1$. However, in the other existing closed-form methods such as \cite{HoXu2004}, at least $N + 2$ sensors are required.

\section{Performance Analysis}\label{sec:performance_analysis}
We express $\boldsymbol{\hat{\theta}}$ and $\Delta \boldsymbol{\hat{\theta}}$, the estimated values in the first and second stages, as $\boldsymbol{\theta} + \Delta \boldsymbol{\theta}$ and $\Delta \boldsymbol{\theta} + \delta \boldsymbol{\theta}$, respectively, which leads to
\begin{equation}
    \tilde{\boldsymbol{\theta}} - \mathbb{E}\!\left\{ \hat{\boldsymbol{\theta}} \right\}
    = \mathbb{E}\!\left\{ \delta\boldsymbol{\theta} \right\} - \delta\boldsymbol{\theta}.
\tag{39}
\end{equation}

By substituting $\mathbf{h}_2 - \mathbf{G}_2\Delta\boldsymbol{\theta}  = \mathbf{B}_2\Delta\mathbf{m}$ from (34) into (36), the error term $\delta \boldsymbol{\theta}$ can be written as
\begin{equation}
    \delta\boldsymbol{\theta} = \Delta \hat{\boldsymbol{\theta}} - \Delta \boldsymbol{\theta}
    = \left( \mathbf{G}_2^{T} \mathbf{W}_2 \mathbf{G}_2 \right)^{-1}
      \mathbf{G}_2^{T} \mathbf{W}_2 \mathbf{B}_2\,\Delta\mathbf{m}.
\tag{40}
\end{equation}

The direct computing of $\mathbb{E} \left\{ \delta \boldsymbol{\theta} \right\}$ is a non-trivial task because there are noise and error terms in both ${\mathbf{G}}_2$ and $\Delta \mathbf{m}$. If the measurement noises are small enough, the error terms in $\hat{\mathbf{G}}_2$ and $\hat{\mathbf{B}}_2$ can be ignored and therefore $\delta \boldsymbol{\theta}$ depends linearly on the measurement noise vector $\Delta \mathbf{m}$ as follows
\begin{equation}
    \delta\boldsymbol{\theta} \approx
    \left( \mathbf{G}_2^{T} \mathbf{W}_2 \mathbf{G}_2 \right)^{-1}
    \mathbf{G}_2^{T} \mathbf{W}_2 \mathbf{B}_2\,\Delta\mathbf{m}.
\tag{41}
\end{equation}

Therefore, it follows from (41) that $\mathbb{E} \left\{ \delta \theta \right\}$ becomes zero and the error covariance matrix of the proposed estimator can be approximately expressed as follows
\begin{equation}
    \mathrm{cov}(\hat{\boldsymbol{\theta}})
    = \mathrm{cov}(\boldsymbol{\delta\theta})
    \approx \left( \mathbf{G}_2^{T} \mathbf{W}_2 \mathbf{G}_2 \right)^{-1}.
\tag{42}
\end{equation}

The CRLB of $\boldsymbol{\theta}$ is found by taking the inverse of the Fisher information matrix. Under the Gaussian measurement noise model, it is simplified to 
\begin{equation}
    \mathrm{CRLB}(\boldsymbol{\theta})
    = \left( \nabla_{\boldsymbol{\theta}}^\mathbf{m}\,^{T}\, \mathbf{Q}_\mathbf{m}^{-1}\,
    \nabla_{\boldsymbol{\theta}}^\mathbf{m} \right)^{-1},
\tag{43}
\end{equation}
where $\nabla_{\theta}^\mathbf{m}$ denotes the partial derivative of the true measurement vector $\mathbf{m}$ with respect to the unknown vector $\boldsymbol{\theta}$ and can be expressed as follows
\begin{equation}
    \nabla_{\boldsymbol{\theta}}^\mathbf{m} =
    \begin{bmatrix}
        \mathbf{C} & \mathbf{0}_{M\times N} \\
        \dot{\mathbf{C}} & \mathbf{C}
    \end{bmatrix},
\tag{44}
\end{equation}
and $\mathbf{C}$ and $\dot{\mathbf{C}}$ are matrices of size $M \times N$ in which the $k$th row is given by
\begin{equation}
    [\mathbf{C}]_{i,:} = \boldsymbol{\rho}_{\mathbf{u}^o, \mathbf{s}_i}^{\,T},
    \qquad
    [\dot{\mathbf{C}}]_{i,:} = \dot{\boldsymbol{\rho}}_{\mathbf{u}^o, \mathbf{s}_i}^{\,T},
\tag{45}
\end{equation}
where $\boldsymbol{\rho}_{\mathbf{a},\mathbf{b}}$ and $\dot{\boldsymbol{\rho}}_{\mathbf{a},\mathbf{b}}$ are given below (7) and (33), respectively, and $k = (i-1) N + j$ for $i = 1, \dots, M$ and $j = 1, \dots, N$.

By substituting the weighting matrix $\mathbf{W}_2$ given by (37) into (42) and defining $\mathbf{G}_3 = \mathbf{B}_2^{-1} \mathbf{G}_2$, it follows that
\begin{equation}
    \mathrm{cov}(\hat{\boldsymbol{\theta}})
    \approx \left( \mathbf{G}_3^{T}\, \mathbf{Q}_\mathbf{m}^{-1}\, \mathbf{G}_3 \right)^{-1}.
\tag{46}
\end{equation}

It is important to note that (46) and the CRLB given by (43) are of the same form. Forming $\mathbf{G}_3$ by doing some straightforward mathematical manipulations, yields
\begin{equation}
    \mathbf{G}_3 \approx \nabla_{\boldsymbol{\theta}}^\mathbf{m}
\tag{47}
\end{equation}
when the following small noise conditions are satisfied:
\[
\text{C1)} \quad |\Delta r_i| \ll \|\mathbf{u}^o - \mathbf{s}_i\|, \quad i = 1, \dots, M
\]
\[
\text{C2)} \quad |\Delta \dot{r}_i| \ll \|\mathbf{u}^o - \mathbf{s}_i\|, \quad i = 1, \dots, M
\]

From (47), we can immediately conclude that
\begin{equation}
    \mathrm{cov}(\hat{\boldsymbol{\theta}}) \approx \mathrm{CRLB}(\boldsymbol{\theta}).
\tag{48}
\end{equation}

\section{Simulation Results}\label{sec:simulation_results}
In this section, we aim to evaluate the performance of the proposed method via different simulation scenarios and corroborate the theoretical results. We consider a sensor network consisting of $M + 1$ sensors in 2-D space ($N = 2$). The TDOA and FDOA measurements are generated by adding to the true values the zero-mean Gaussian noises with a covariance matrix $Q = \sigma^2 \left( \mathbf{1}_{M \times M} + \mathbf{I}_M \right)/2$,
where $\sigma^2$ is the TDOA and FDOA variance. To evaluate performance of the proposed method, we have utilized the Root Mean Square Error (RMSE) criterion, which can be computed via a Monte Carlo simulation as $\text{MSE}(\mathbf{u}) = \sqrt{\frac{1}{L} \sum_{l=1}^{L} \|\mathbf{u}^{(l)} - \mathbf{u}^o\|^2}$,
where $\mathbf{u}^{(l)}$ denotes the estimate of $\mathbf{u}^o$ at the $l$-th ensemble run. The number of Monte Carlo runs in all scenarios, $L$, is 5000.

\subsection*{A. Scenario 1: 2-D Localization with Minimum Sensors in Challenging Geometry}

\begin{figure}[!t]
\centering
\subfloat[Position RMSE vs.\ noise variance]{%
  \includegraphics[width=\columnwidth]{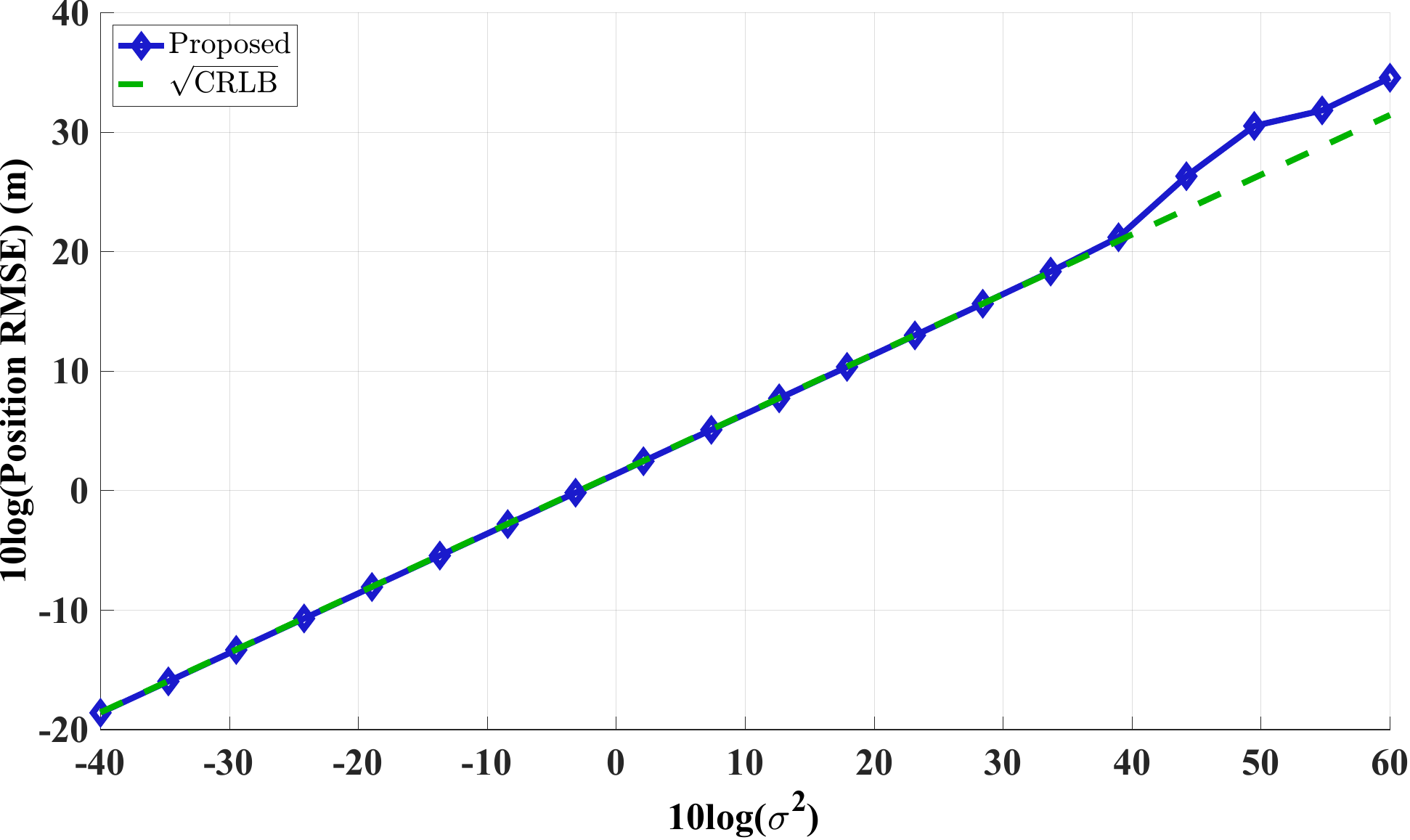}%
  \label{fig:pos_rmse_n2}}
\vspace{2mm}
\subfloat[Velocity RMSE vs.\ noise variance]{%
  \includegraphics[width=\columnwidth]{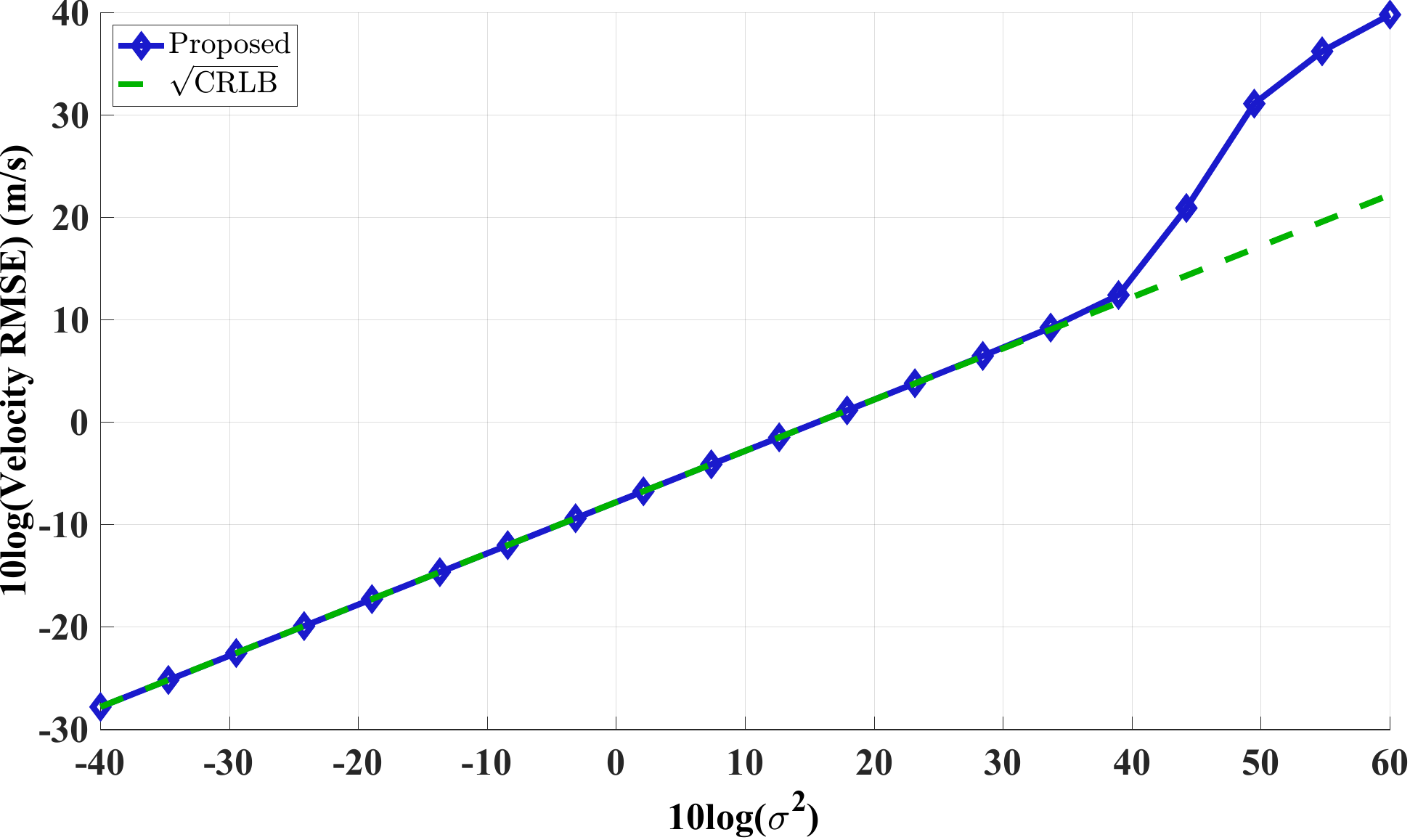}%
  \label{fig:vel_rmse_n2}}
\caption{RMSE performance of the proposed method compared with the CRLB in the first scenario.}
\label{fig:rmse_n2}
\end{figure}

We first evaluate the performance when only three sensors---the theoretical minimum for 2-D localization---are available, and placed in a near‐colinear configuration to stress observability. The sensor positions and velocities are tabulated in Table~\ref{tab:sensors_state_2d} (indices 0-2), and source is at \(\mathbf{u}^\circ = [400,\;200]^T\) m with velocity \(\mathbf{\dot{u}}^\circ = [20,\;10]^T\) m/s. In Fig.~\ref{fig:rmse_n2}, we plot \(\mathrm{RMSE}(\mathbf{u})\) and \(\mathrm{RMSE}(\mathbf{\dot{u}})\) as a function of \(10 \log(\sigma^2)\), where noise variance is \(\sigma^2\in[10^{-4},\,10^6]\) m\(^2\). As opposed to other methods, like (i) Noroozi's TSWLS \cite{NorooziOveis2018}, (ii) Ho's TSWLS \cite{HoXu2004}, (iii) ICWLS \cite{QuXieTan2017}, and (iv) SDP \cite{WangLiAnsari2013}, the proposed estimator can estimate the position and velocity of the source with the minimum number of sensors required for this purpose. Other estimators cannot estimate with this number of sensors and need at least one more sensor.

\begin{table}[t]
\centering
\caption{True 2-D Position (m) and Velocity (m/s) of Sensors}
\label{tab:sensors_state_2d}
\renewcommand{\arraystretch}{1.2}
\setlength{\tabcolsep}{12pt}
\begin{tabular}{c r r r r}
\hline\hline
\textbf{Sensor} $i$ & $x_i$ & $y_i$ & $\dot{x}_i$ & $\dot{y}_i$ \\
\hline
0 (ref.) &   50 &   50 &  20 &  30 \\
1        & 1000 & 1000 & -10 & -10 \\
2        &  200 &  800 &  50 &  20 \\
3        &  500 &  100 & -30 &  10 \\
\hline\hline
\end{tabular}
\end{table}

\subsection*{B. Scenario 2: 2-D Localization with Four Sensors and Algorithm Comparison}
\begin{figure}[!t]
\centering
\subfloat[Position RMSE vs.\ noise variance]{%
  \includegraphics[width=\columnwidth]{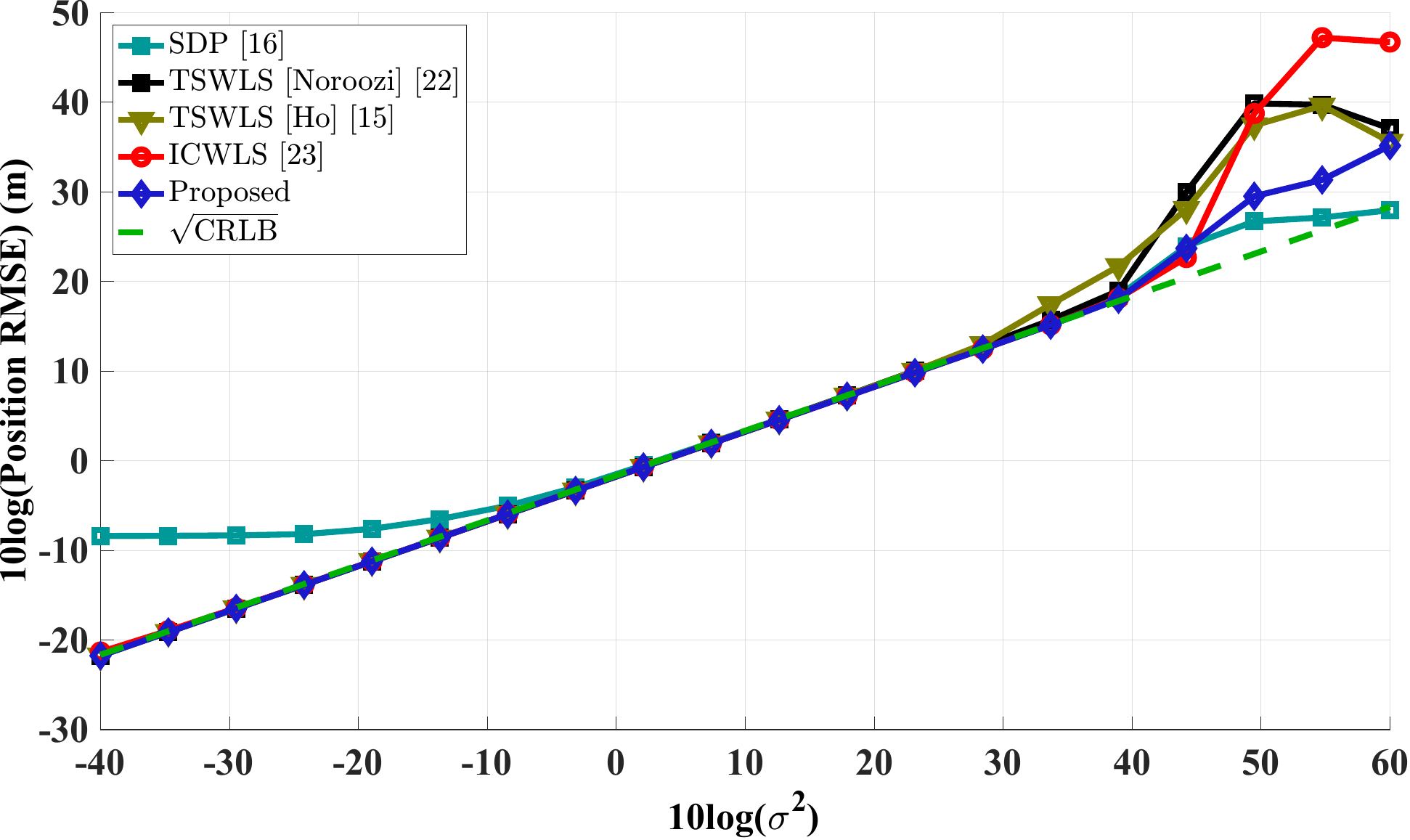}%
  \label{fig:pos_rmse_n3}}
\vspace{2mm}
\subfloat[Velocity RMSE vs.\ noise variance]{%
  \includegraphics[width=\columnwidth]{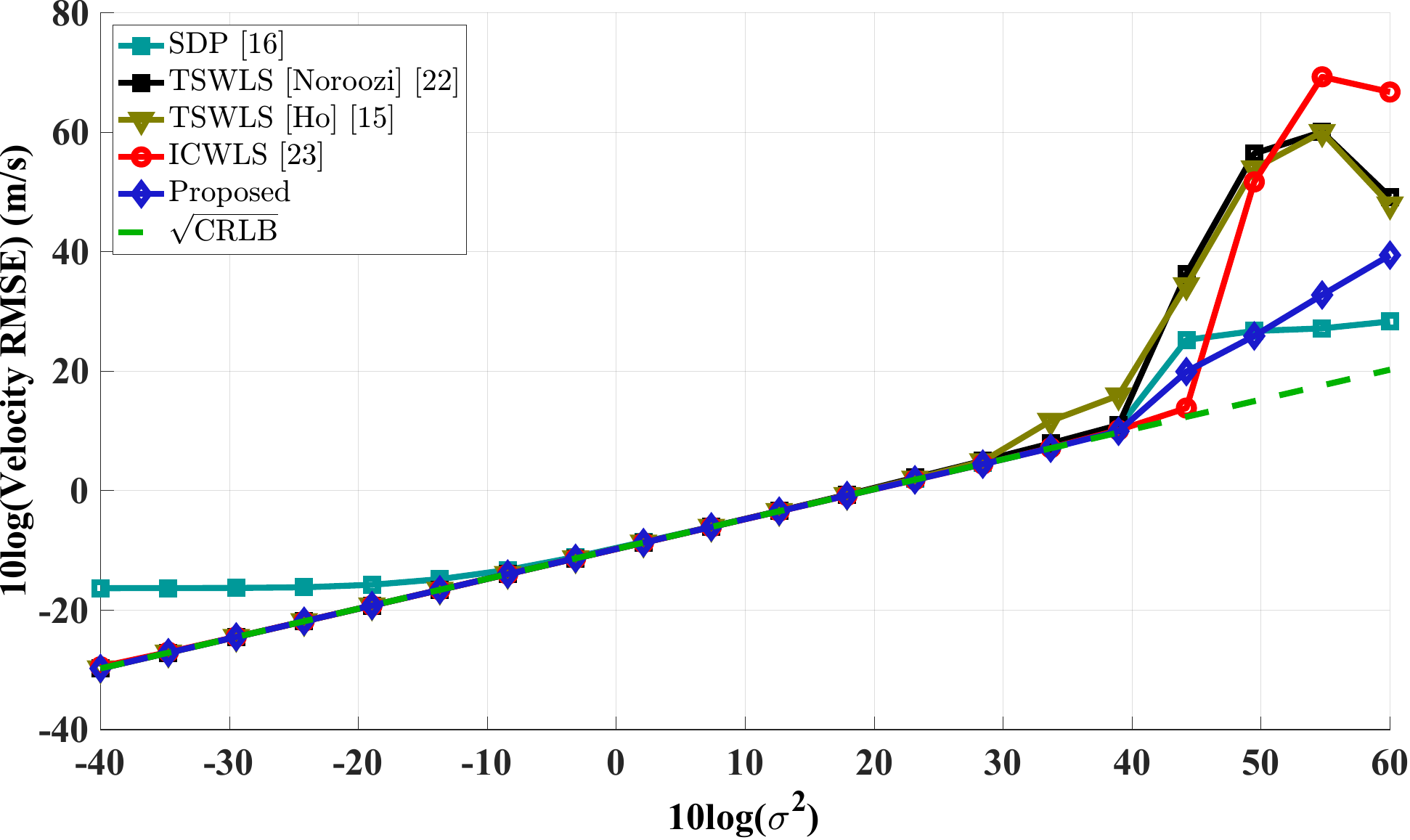}%
  \label{fig:vel_rmse_n3}}
\caption{RMSE performance of the proposed method compared with the CRLB and other state-of-the-art estimators in the second scenario.}
\label{fig:rmse_n3}
\end{figure}

Next, we consider four sensors (one above the minimum required) as listed in Table~\ref{tab:sensors_state_2d} (indices 0-3). The source is at \(\mathbf{u}^\circ=[400,\;200]^T\) m with velocity \(\mathbf{\dot{u}}^\circ = [20,\;10]^T\) m/s. We compare our method against (i) Noroozi's TSWLS \cite{NorooziOveis2018}, (ii) Ho's TSWLS \cite{HoXu2004}, (iii) ICWLS \cite{QuXieTan2017}, and (iv) SDP \cite{WangLiAnsari2013}. In Fig.~\ref{fig:rmse_n3}, we show RMSE of position and velocity versus \(10\log(\sigma^2)\). The proposed estimator consistently outperforms closed-form alternatives, achieving near-CRLB-level error across the entire SNR range and performing close to the more complex SDP method, especially at higher noise levels where other methods degrade significantly.

\subsection*{C. Scenario 3: Random Sensor and Source Deployment}

\begin{figure}[!t]
\centering
\includegraphics[width=\columnwidth]{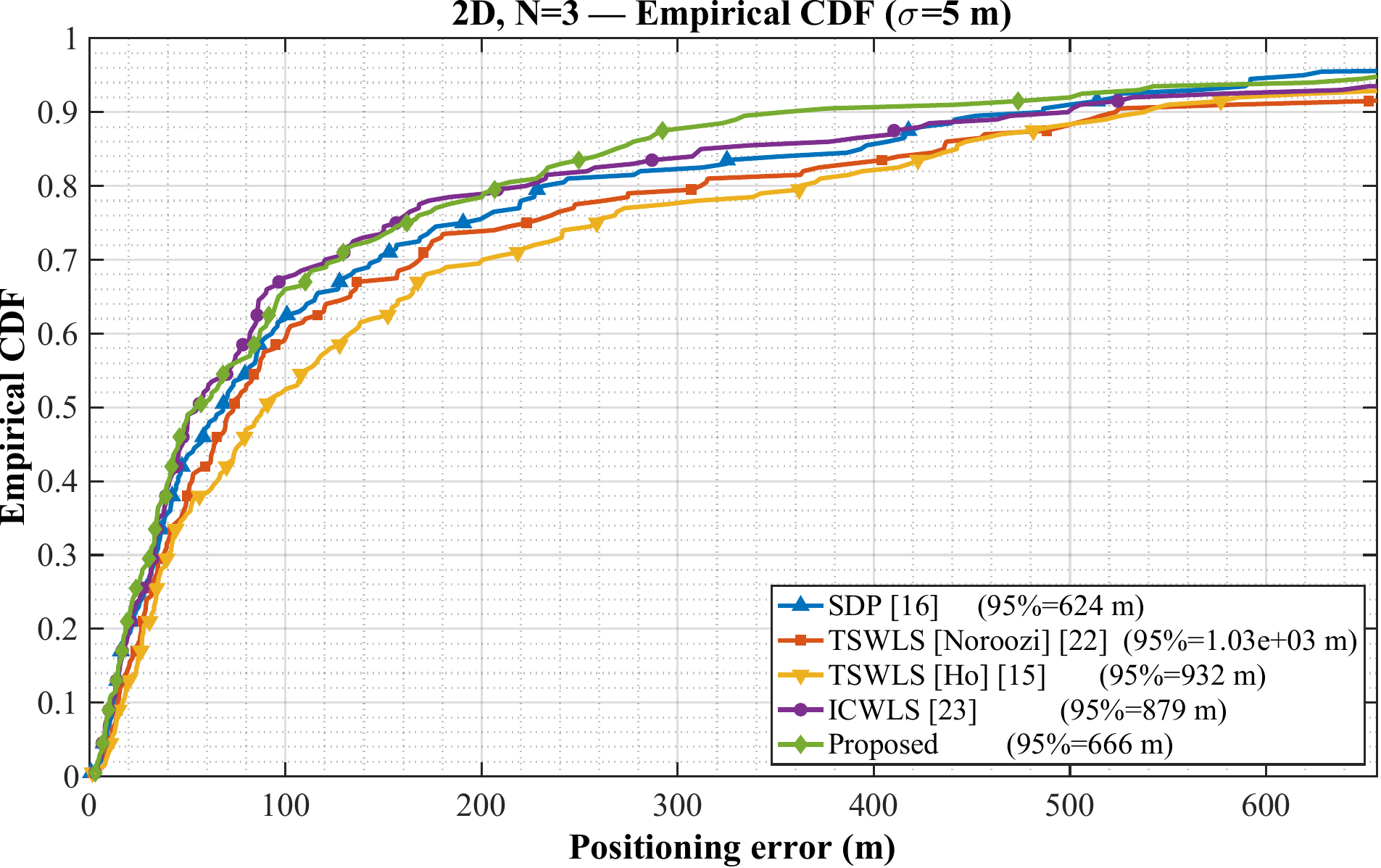} % or .pdf if you export it
\caption{2D, $N\!=\!3$ — Empirical CDF of positioning error for $\sigma=5$ m.
The legend reports the $95\%$ error for each method.}
\label{fig:cdf_2d_n3}
\end{figure}

To assess robustness to geometry, we uniformly place six sensors in the plane, \(\mathbf{s}_i\sim\mathcal{U}([0,1000]^2)\), and draw the source uniformly from the same square. For each of 200 Monte Carlo trials, we generate TDOA/FDOA measurements with a level noise of \(\sigma=5\) m. Then, compute the empirical CDF of the 2-D positioning error. Fig.~\ref{fig:cdf_2d_n3} overlays the CDFs of our method, Noroozi's TSWLS \cite{NorooziOveis2018}, Ho's TSWLS \cite{HoXu2004}, ICWLS \cite{QuXieTan2017}, and SDP \cite{WangLiAnsari2013}. As depicted, even under random, potentially ill-conditioned geometries, the proposed approach yields a performance improvement compared with other closed-form methods.

\begin{table}[t]
\centering
\caption{True 3-D Position (m) and Velocity (m/s) of Sensors}
\label{tab:sensors_state_3d}
\renewcommand{\arraystretch}{1.2}
\setlength{\tabcolsep}{8pt}
\begin{tabular}{c r r r r r r}
\hline\hline
\textbf{Sensor} $i$ & $x_i$ & $y_i$ & $z_i$ & $\dot{x}_i$ & $\dot{y}_i$ & $\dot{z}_i$ \\
\hline
0 (ref.) &  300 &  100 &  150 &  30 & -20 &  20 \\
1        &  400 &  150 &  100 & -30 &  10 &  20 \\
2        &  300 &  500 &  200 &  10 & -20 &  10 \\
3        &  350 &  200 &  100 &  10 &  20 &  30 \\
4        & -100 & -100 & -100 & -20 &  10 &  10 \\
\hline\hline
\end{tabular}
\end{table}

\begin{figure}[!t]
\centering
\subfloat[Position RMSE vs.\ noise variance (3D, $N=4$).]{%
  \includegraphics[width=\columnwidth]{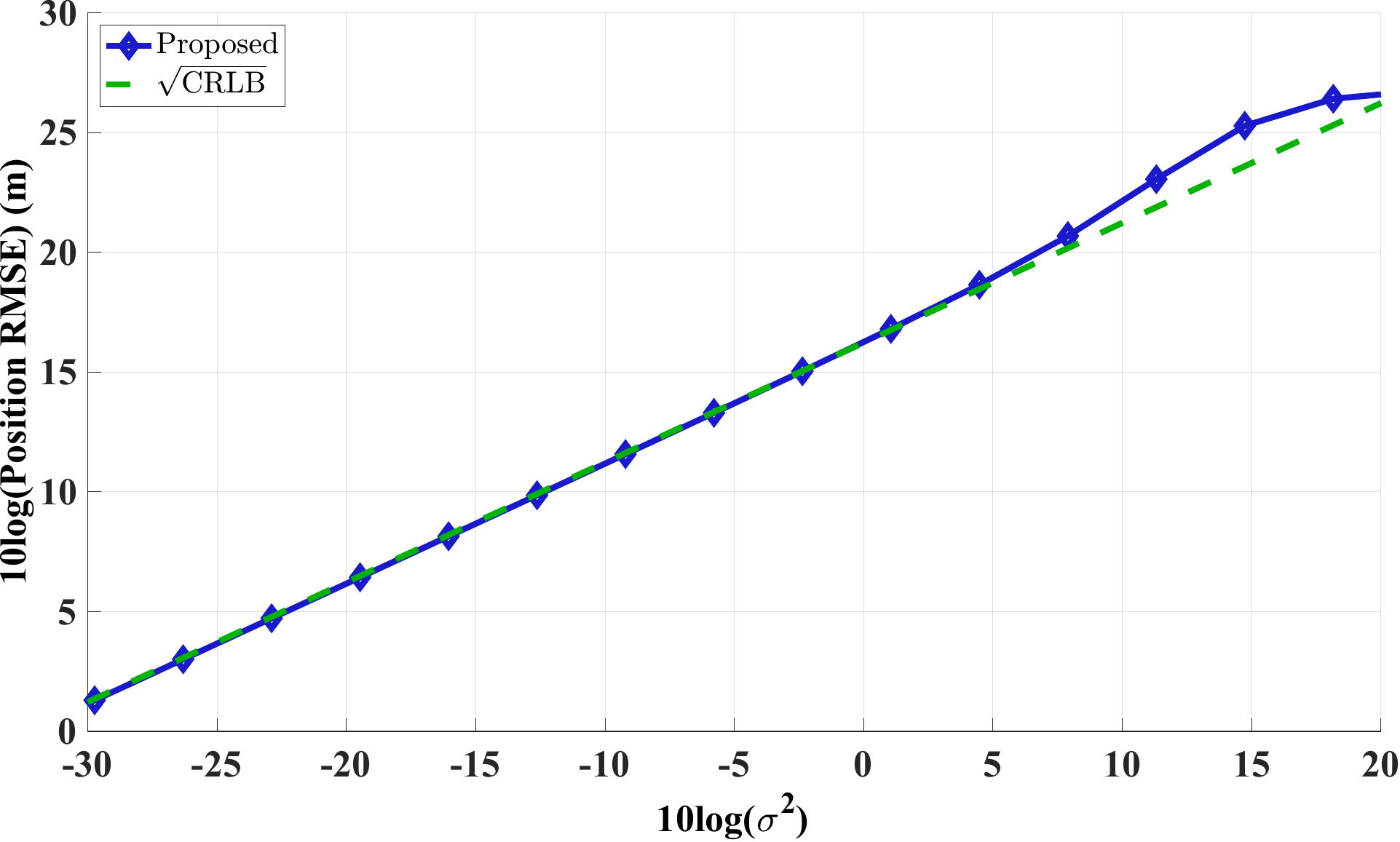}%
  \label{fig:pos_rmse_3d_n4}}
\vspace{2mm}
\subfloat[Velocity RMSE vs.\ noise variance (3D, $N=4$).]{%
  \includegraphics[width=\columnwidth]{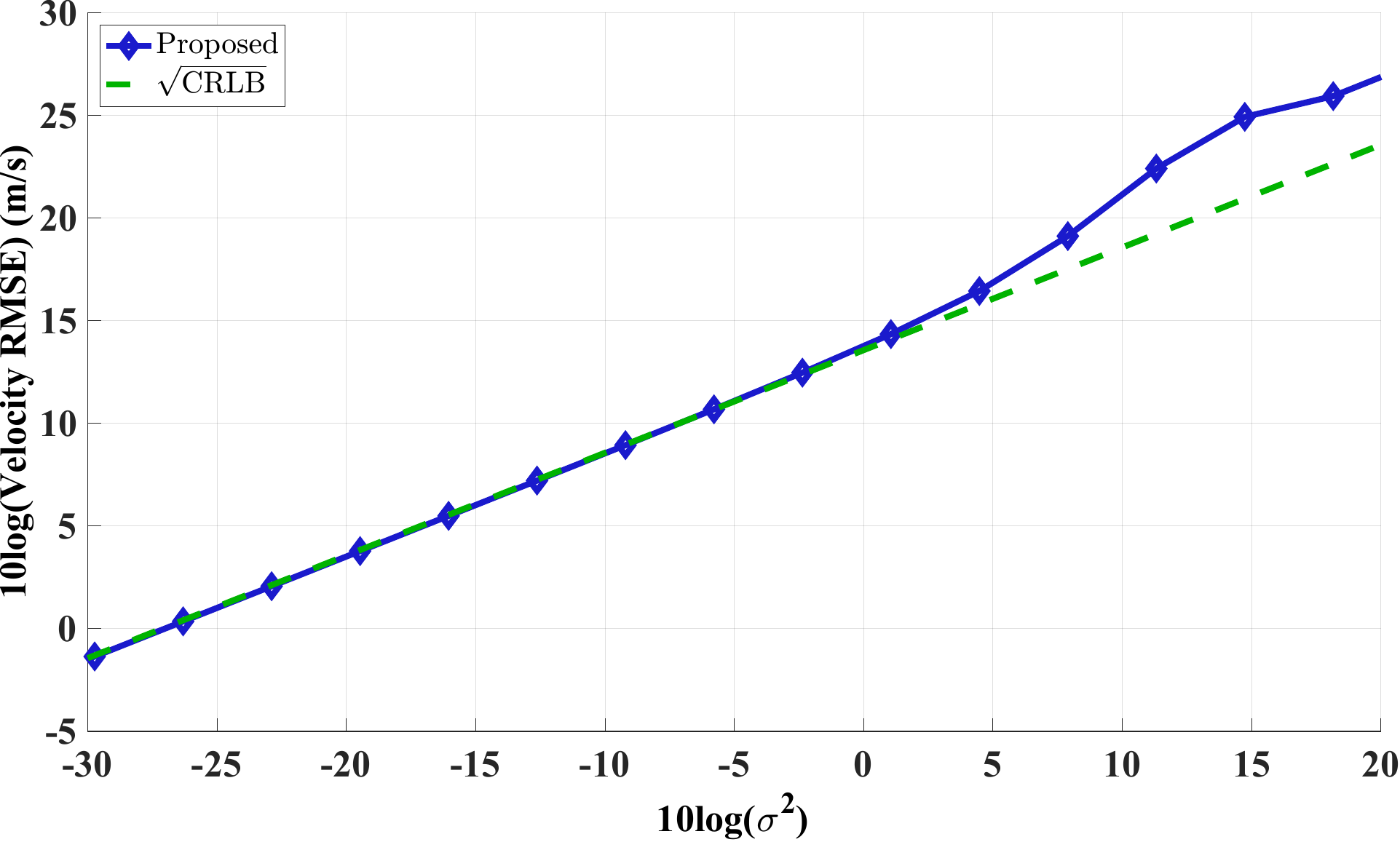}%
  \label{fig:vel_rmse_3d_n4}}
\caption{3-D RMSE performance of the proposed method compared with the CRLB in the fourth scenario.}
\label{fig:rmse_3d_n4}
\end{figure}

\subsection*{D. Scenario 4: 3-D Localization with Minimum Sensors}
To demonstrate the performance of the proposed method, we design another scenario for localizing in 3D. Consequently, we use the minimum number of sensors required for localizing in 3D, i.e., four sensors. The sensors' position and velocity are the first four sensors of the Table~\ref {tab:sensors_state_3d}. Also, the source is at \(\mathbf{u}^\circ=[600,\,650,\,550]^T\) m with velocity \(\mathbf{\dot{u}}^\circ = [-20,\,15,\,40]^T\) m/s. In Fig.~\ref{fig:rmse_3d_n4}, we plot \(\mathrm{RMSE}(\mathbf{u})\) and \(\mathrm{RMSE}(\mathbf{\dot{u}})\) versus \(10\log(\sigma^2)\). The results confirm that our estimator attains the 3D CRLB with only four sensors, whereas existing closed‐form methods require at least five sensors to locate the source.

\subsection*{E. Scenario 5: 3-D Localization with Five Sensors}

\begin{figure}[!t]
\centering
\subfloat[Position RMSE vs.\ noise variance (3D, $N=5$).]{%
  \includegraphics[width=\columnwidth]{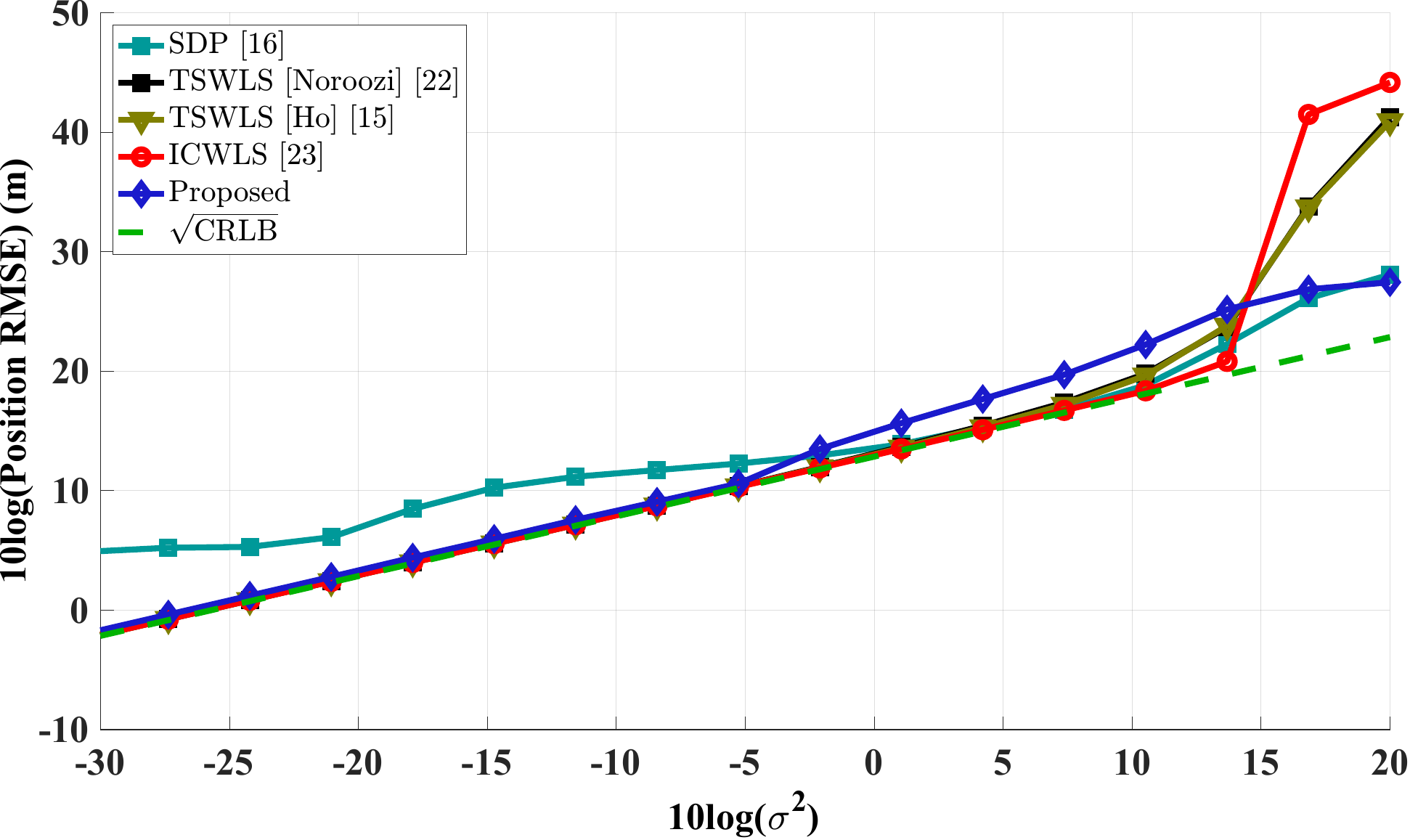}%
  \label{fig:pos_rmse_3d_n5}}
\vspace{2mm}
\subfloat[Velocity RMSE vs.\ noise variance (3D, $N=5$).]{%
  \includegraphics[width=\columnwidth]{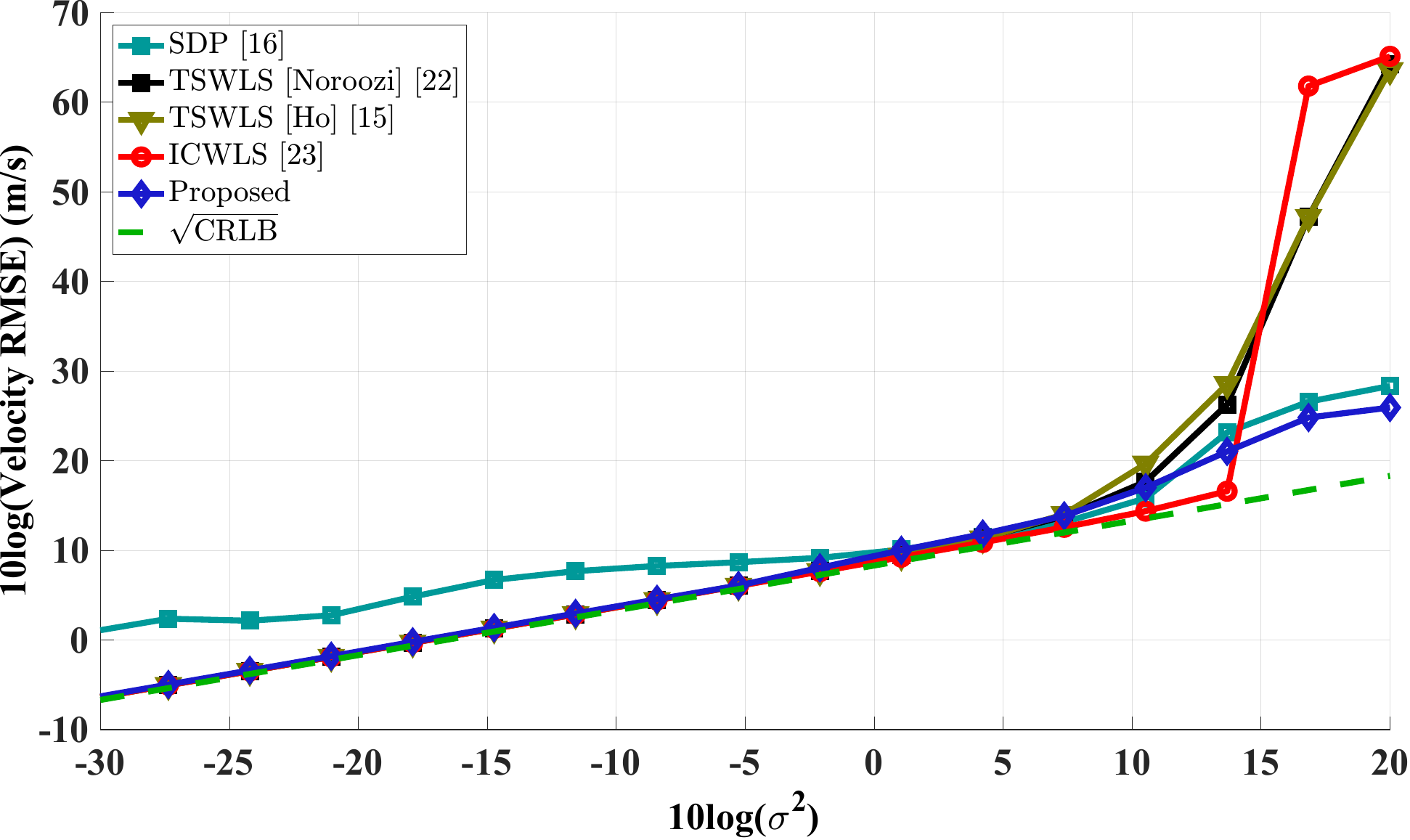}%
  \label{fig:vel_rmse_3d_n5}}
\caption{3-D RMSE performance of the proposed method compared with the CRLB and other state-of-the-art estimators in the fifth scenario.}
\label{fig:rmse_3d_n5}
\end{figure}

To demonstrate the performance of the proposed method in higher dimensions and with other state-of-the-art estimators, we design another scenario for localizing in 3D using five sensors, i.e., more than the minimum sensors needed. The sensors' position and velocity are listed in the Table~\ref {tab:sensors_state_3d} (indices 0-4). In addition, the source is at \(\mathbf{u}^\circ=[600,\,650,\,550]^T\) m with velocity \(\mathbf{\dot{u}}^\circ = [-20,\,15,\,40]^T\) m/s. In Fig.~\ref{fig:rmse_3d_n5}, we plot \(\mathrm{RMSE}(\mathbf{u})\) and \(\mathrm{RMSE}(\mathbf{\dot{u}})\) versus \(10\log(\sigma^2)\). The results show that the proposed estimator can work properly in higher dimensions and perform well in relatively higher noise than other methods.

\subsection*{F. Scenario 6: 2-D Localization comparing with different number of sensors}
\begin{figure}[!t]
\centering
\subfloat[Position RMSE vs.\ total number of sensors (log scale).]{%
  \includegraphics[width=\columnwidth]{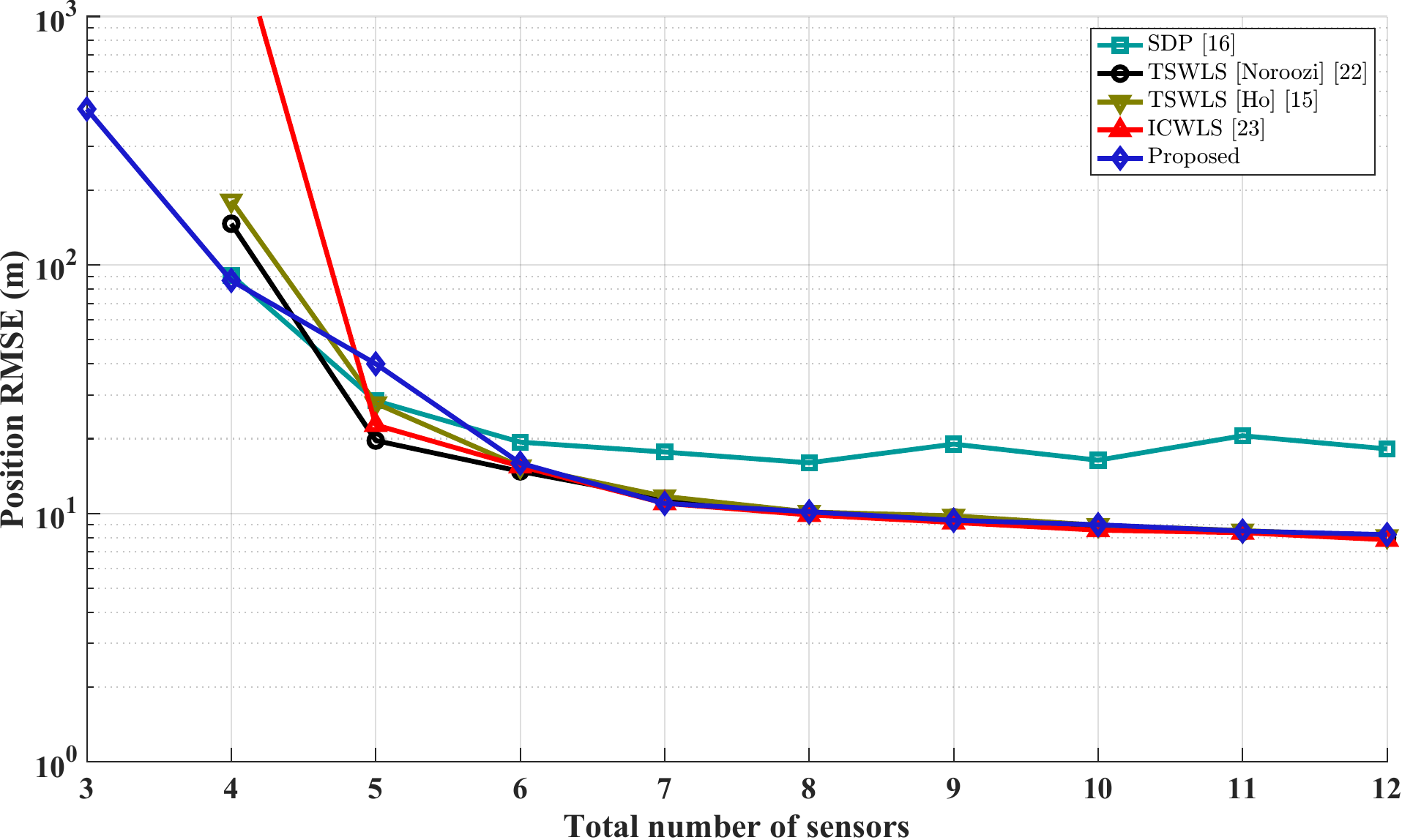}%
  \label{fig:pos_rmse_vs_M}}
\vspace{2mm}
\subfloat[Velocity RMSE vs.\ total number of sensors.]{%
  \includegraphics[width=\columnwidth]{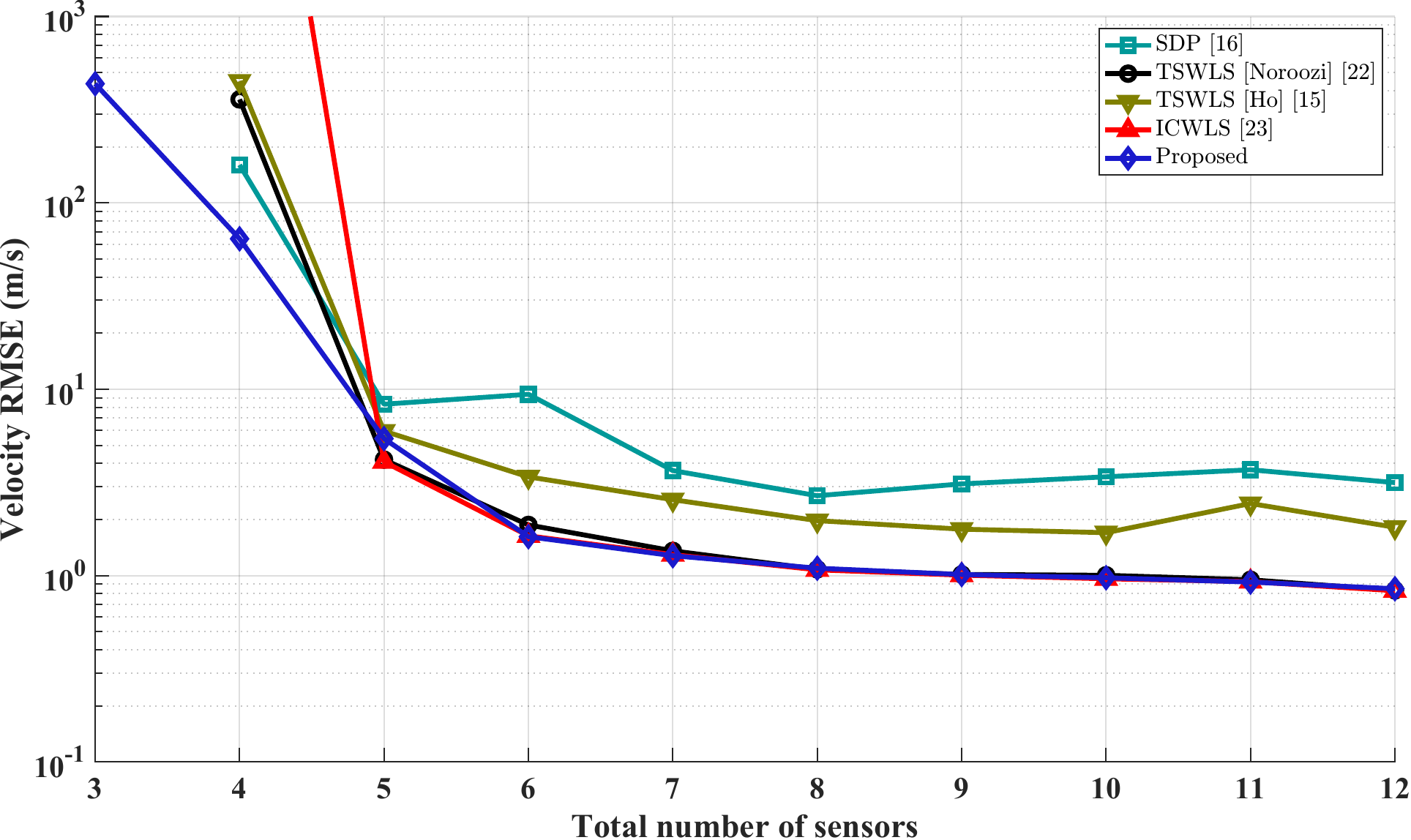}%
  \label{fig:vel_rmse_vs_M}}
\caption{Effect of sensor count on RMSE (random geometry per run).}
\label{fig:rmse_vs_M}
\end{figure}

In this scenario, we investigate how the estimation accuracy of different TDOA-FDOA localization methods improves as the number of sensors increases. The setup is in a 2D plane where the source  position is \(\mathbf{u}^\circ=[550,\,450]^T\) m with velocity \(\mathbf{\dot{u}}^\circ = [-10,\,5]^T\) m/s. Sensors are randomly deployed in a cubic area \(\mathbf{s}_i\sim\mathcal{U}([0,1000]^2)\). The results are depicted in the Fig~\ref{fig:rmse_vs_M}. In each test, a Monte Carlo run is performed 500 times to capture a variety of geometries. Measurement noise is modeled using covariance matrices scaled by a fixed noise variance level (\(\sigma^2 = 25\)). Monte Carlo averaging over multiple random sensor deployments ensures statistically meaningful results. The performance metric is the RMSE of both position and velocity estimates, evaluated as a function of the total number of sensors (from 3 to 12).
As the figures show, the proposed method and ICWLS offer more robust performance across different geometries than other approaches. Crucially, the proposed method holds a key advantage: it is the only one capable of locating a source with only three sensors.

\section{Conclusion}\label{sec:conclusion}

We have presented a novel two‐stage algebraic estimator for moving‐source localization that fuses TDOA and FDOA measurements using only $N+1$ sensors in $N$-dimensional space.  In the first stage, pseudo‐linear equations with range and range‐rate nuisance parameters are solved in closed form by weighted least squares and Sylvester’s resultant, yielding explicit quartic solutions for both range and range‐rate.  A second‐stage linear refinement then corrects residual errors to produce position and velocity estimates that attain the CRLB under mild Gaussian noise.  Monte Carlo simulations in 2‐D and 3‐D validate that our method matches CRLB accuracy and outperforms existing two‐stage and iterative schemes, particularly in low‐sensor or high‐noise regimes.  By operating at the theoretical minimal‐sensor bound, the proposed estimator is well suited to size‐, weight‐, and power‐constrained applications.

% -------------------------
% References
% -------------------------
\bibliographystyle{IEEEtran}

\end{document}